\documentclass[10pt]{amsart}

\usepackage{mathrsfs}

\usepackage{xpatch}
\xpatchcmd{\paragraph}{\normalfont}{{\normalfont\bfseries}}{}{}

\usepackage{color}

\newcommand{\CC}{\mathbb{C}}

\newcommand{\ben}{\begin{eqnarray}}
\newcommand{\een}{\end{eqnarray}}

\newcommand{\non}{\nonumber}

\begin{document}

\title{Scalar product for the XXZ spin chain with general integrable boundaries}

\dedicatory{Dedicated to the memory of Omar Foda}

\author{Samuel Belliard$^{*}$}

\address{$^*$ Institut Denis-Poisson CNRS/UMR 7013 - Universit\'e de Tours - Universit\'e d'Orl\'eans
Parc de Grammont, 37200 Tours, FRANCE}

\email{samuel.belliard@gmail.com}

\author{Rodrigo A. Pimenta$^{**}$}
\address{$^{**}$Departamento de F\'isica, Universidade Federal de Lavras,
          Caixa Postal 3037, 37200-000, Lavras, MG, BRAZIL}
\address{Instituto de F\'isica de S\~ao Carlos, Universidade de S\~ao Paulo, Caixa Postal 369, 13.560-590, S\~ao Carlos, SP, BRAZIL}

\email{pimenta@ifsc.usp.br}

\author{Nikita A. Slavnov$^{***}$}

\address{Steklov Mathematical Institute of Russian Academy of Sciences, 8 Gubkina str., Moscow, 119991,  RUSSIA}

\email{nslavnov@mi-ras.ru}

\begin{abstract}
We calculate the scalar product of Bethe states of the XXZ spin-$\frac{1}{2}$ chain with general integrable boundary conditions. The off-shell equations satisfied by the transfer matrix and the off-shell Bethe vectors
allow one to derive a linear system for the scalar product of off-shell and on-shell Bethe states. We show that this linear system can be solved
in terms of a compact determinant formula that involves the Jacobian of the transfer
matrix eigenvalue and certain q-Pochhammer polynomials of the boundary couplings.
\end{abstract}

\maketitle


\paragraph{Introduction.}


The anisotropic Heisenberg  XXZ spin chain is a paradigmatic model of interacting many body quantum system. It is an integrable model within the quantum inverse scattering method \cite{SFT,Skly88}. In this paper, we consider the XXZ spin chain on the segment with general integrable boundaries given by the Hamiltonian
\ben\label{H}
&&H=
\epsilon\, \sigma^{3}_1 +  \kappa^-\,\sigma^{-}_1  +  \kappa^+\,\sigma^{+}_1   +
\sum_{k=1}^{N-1}\Big(
\sigma^{x}_{k}\otimes \sigma^{x}_{k+1}+\sigma^{y}_{k}\otimes \sigma^{y}_{k+1} +
\Delta\sigma^{3}_{k}\otimes \sigma^{3}_{k+1}\Big)  +
\nu\,  \sigma^{3}_N +\tau^-\,  \sigma^{-}_N+\tau^+\,  \sigma^{+}_N,
\een
where $N$ is the length of the chain, $\sigma_i^{\pm,x,y,3}$ are the standard Pauli matrices acting non-trivially in the $i$-th site of the quantum space $\mathcal{H}= \otimes_{i=1}^N \CC^2$. The anisotropy parameter is $\Delta=\frac{q+q^{-1}}{2}$ with generic $q$. Left $\{ \epsilon,  \kappa^\pm\}$ and right
$\{ \nu,  \tau^\pm\}$ boundary couplings are also generic. The aim of this
paper is to compute the scalar product of on-shell and off-shell Bethe
states of the Hamiltonian (\ref{H}).

The building blocks of the quantum inverse scattering method are the R-matrix, which solves the Yang--Baxter equations, and the K-matrices, which solve the reflection equation \cite{GZ, deVega}. They can be used to construct commuting transfer matrices, which contain the XXZ Hamiltonian with arbitrary  boundary couplings (\ref{H}) as a conserved charge. Solving the spectral problem of the transfer matrix is the primordial aim, followed by the computation of scalar products, form factors and correlation functions (see, {\it e.g.}, \cite{KBI}).
One of the most powerful methods available to handle such problems is the algebraic Bethe ansatz:
it allows the construction of off-shell Bethe states of the transfer matrix as well as certain scalar products in terms of
compact formulas (see \cite{Srew} for a recent review).

Let us recall that the K-matrices are associated with boundary fields at each end of the spin chain.
The boundaries in general can break the $U(1)$ symmetry of the bulk Hamiltonian, and
it happens when the K-matrices have a general non-diagonal form.
This breaking prevents the application of the standard Bethe ansatz technique to analyze the
spectral problem of the transfer matrix, unless some constraints on the boundary couplings are imposed (see e.g. \cite{gauge,YanZ07,TQ1,TQ2,BCR12,PL13,AMS})\footnote{We do not intend
to give an exhaustive list of references on this previous contributions here, and we refer to paper \cite{B15} for more details on the solution of this problem.}.
In particular, the construction of off-shell states of the transfer matrix in the generic boundary case
requires using the modified algebraic Bethe ansatz \cite{BC13,B15,Cra14,BelPim15,ABGP15}. This method is based on constructing a modified creation operator that satisfies a certain new off-shell relation, which adds an extra term in the transfer matrix eigenvalue.  This extra term in the eigenvalue expression
is called an inhomogeneous term, firstly discovered \cite{CYSW1} within the off-diagonal Bethe ansatz method (ODBA) \cite{ODBAbook} (see also \cite{Nepo13,KMN14}).
We note that on-shell Bethe states have been constructed in the ODBA framework \cite{CYSW5}. We also recall that study of $U(1)$ breaking boundaries in the XXZ chain has been considered from different and complementary perspectives, including the
q-Onsager method \cite{B04,BK07}, the quantum separation of variables \cite{FKN,KMN14,KMNT} and the Q-operator method \cite{LP14}.

Modified algebraic Bethe ansatz was further developed by considering scalar products of Bethe vectors. In particular, considering isotropic XXX spin-$1/2$ chain with non-diagonal twist and on the segment with general integrable boundary conditions, compact formulas for the scalar product of on-shell and off-shell Bethe vector have been conjectured \cite{BP15scalar1,BP15scalar2}.
These formulas involve the Jacobian of the transfer matrix eigenvalue, similarly to diagonal boundary cases, but contain a modified
factor, which is due to the broken $U(1)$ symmetry.
More recently, the case of generic twisted boundary conditions has been proved in \cite{BS19,BS19det}.

It was shown in \cite{BS19det,XYZscalar} that the scalar product of on-shell and off-shell Bethe vector satisfies a homogeneous system of linear equations that follow
directly from the transfer matrix action on the off-shell Bethe vector.  Another key ingredient of the method developed in \cite{BS19det} is
the asymptotic behavior of the off-shell Bethe vector when its arguments go to infinity. In this paper, we develop this program for the XXZ chain with general integrable boundary conditions and obtain a compact formula  (\ref{scalar}), paving the way to further studies of form factors and correlation functions of the model. The formula (\ref{scalar}) generalizes the conjecture of \cite{BP15scalar2} to the q-deformed case. It also generalizes previous computations of scalar products \cite{scalarXXX,KKMNST,Yang:2010ma} which were
considered for diagonal or constrained boundary parameters cases.
\\


\paragraph{R-matrix and K-matrices.}  We recall the basic ingredients of the quantum inverse scattering method. The main object is the R-matrix, given by
\ben\label{Rmatrix}
R(u)=
\frac{1}{q-q^{-1}}\left(
\begin{array}{cccc}
qu-q^{-1}u^{-1} & 0 & 0 & 0 \\
 0 & u-u^{-1} & q-q^{-1} & 0 \\
 0 & q-q^{-1} & u-u^{-1} & 0 \\
 0 & 0 & 0 & qu-q^{-1}u^{-1} \\
\end{array}
\right)\,.
\een
The R-matrix acts on the space $\mathbb{C}^2\otimes \mathbb{C}^2$ and is a solution of the Yang--Baxter
equation in $\mathbb{C}^2\otimes \mathbb{C}^2 \otimes \mathbb{C}^2$
\ben
R_{12}(u/v)R_{13}(u)R_{23}(v)=R_{23}(v)R_{13}(u)R_{12}(u/v)\,.
\een
Here the indices $ij$ in $R_{ij}(u)$ indicate the vector spaces where it acts non-trivially. Next, in order to consider the boundary fields, one introduces the reflection matrices
\ben
&&K^{-}(u) =\left(
\begin{array}{cc}
 \nu_- u+\nu_+u^{-1}  & \tau^2(u^2-u^{-2}) \\
\tilde\tau^2(u^2-u^{-2}) &   \nu_- u^{-1}+\nu_+u  \\
\end{array}
\right)\,,\\&&	  K^{+}(u)=
\left(
\begin{array}{cc}
\epsilon_+ q u+\epsilon_-q^{-1}u^{-1}  & \tilde\kappa^2(q^2u^2-q^{-2}u^{-2}) \\
\kappa^2 (q^2u^2-q^{-2}u^{-2}) &  \epsilon_- q u+\epsilon_+q^{-1}u^{-1} \\
\end{array}
\right),
\een
where $\{\nu_{\pm},\tau,\tilde\tau,\epsilon_{\pm},\kappa,\tilde\kappa\}$ are arbitrary  parameters \cite{GZ,deVega}. They are the most general solutions of
the reflection equation and of the dual reflection equation \cite{Skly88}, which hold in $\CC^2\otimes\CC^2$:
\ben \label{RE}
 &&R_{12}(u/v)K^-_1(u)R_{12}(uv)K^-_2(v)=
K^-_2(v)R_{12}(uv)K^-_1(u)R_{12}(u/v),\\
\label{DRE}
&&R_{12}(v/u)K^+_1(u)R_{12}(q^{-2}u^{-1}v^{-1})K^+_2(v)=
K^+_2(v)R_{12}(q^{-2}u^{-1}v^{-1})K^+_1(u)R_{12}(v/u).
\een
From the $R$ and $K^{\pm}$ matrices, one can define the so-called double-row transfer matrix
\ben\label{tra}
t(u)=\operatorname{Tr}_a (K_a^+(u) T_a(u)K_a^{-}(u)\hat T_a(u)),
\een
where $a$ denotes an auxiliary $\CC^2$ vector space, and
\ben
T_a(u) = R_{a1}(u/x_1)R_{a2}(u/x_2)\cdots R_{aN}(u/x_N)\,,
\een
\ben
\hat T_a(u) = R_{aN}(u x_N)R_{aN-1}(ux_{N-1})\cdots R_{a1}(ux_1)\,,
\een
are called
single-row monodromy matrices. They depend on arbitrary parameters  $x_i$ called inhomogeneities. We denote
\ben\label{KO}
K_a(u) = T_a(u)K_a^{-}(u)\hat T_a(u) = \left(\begin{array}{cc}
       \mathscr{A}(u) & \mathscr{B}(u)\\
       \mathscr{C}(u) & \mathscr{D}(u)+\frac{q-q^{-1}}{qu^2-q^{-1}u^{-2}}\mathscr{A}(u)
      \end{array}
\right)_a,
\een
where $\mathscr{A}(u),\mathscr{B}(u),\mathscr{C}(u),\mathscr{D}(u)$ act on $\otimes_{i=1}^N \mathbb{C}^{2}$ and are called double-row monodromy operators.
It turns out that the transfer matrix (\ref{tra})
enjoys the fundamental property
\ben
\left[t(u),t(v)\right]=0\,,
\een
for arbitrary $u$ and $v$. Thus, the transfer matrix (\ref{tra}) is a generating function of conserved charges of the model that allows one to reconstruct the Hamiltonian (\ref{H}) through the relation
\ben\label{Htotr}
H=\frac{q-q^{-1}}{2} \frac{d}{du}\ln(t(u))
\Big|_{u=1, w_i=1}-
\left(
N~\frac{q+q^{-1}}{2}+\frac{(q-q^{-1})^2}{2(q+q^{-1})}
\right).
\een
In terms of the boundary parameters of the $K^{\pm}$-matrices,
the couplings of the Hamiltonian (\ref{H}) are expressed as
\ben\label{parKH1}
&&\epsilon=\frac{(q-q^{-1})}{2}\frac{(\epsilon_+ - \epsilon_-)}{(\epsilon_+ + \epsilon_-)}
,\quad
  \kappa^-=\frac{2(q-q^{-1})}{(\epsilon_+ + \epsilon_-)}\kappa^2,
  \quad   \kappa^+=\frac{2(q-q^{-1})}{(\epsilon_+ + \epsilon_-)}\tilde \kappa^2,\\
  \label{parKH2}
&& \nu=  \frac{(q-q^{-1})}{2}\frac{(\nu_- - \nu_+)}{(\nu_+ +\nu_-)},\quad
 \tau^-= \frac{2(q-q^{-1})}{(\nu_+ + \nu_-)}\tilde \tau^2,\quad
 \tau^+= \frac{2(q-q^{-1})}{(\nu_+ + \nu_-)}\tau^2.
 \een
For convenience, we will use the following parametrization for the boundary parameters:
\ben
&&\nu_-=i\tilde\tau\tau\big(\mu/\tilde \mu+\tilde \mu/\mu\big),
 \quad \nu_+=i\tilde\tau\tau\big(\mu\tilde\mu+1/( \mu\tilde \mu)\big)\,,
\nonumber\\
&&
\epsilon_-=i\tilde\kappa\kappa\big(\xi/\tilde \xi+\tilde\xi/ \xi\big),
\quad \epsilon_+=i\tilde\kappa\kappa\big(\xi \tilde \xi+1/(\tilde\xi \xi)\big).\label{paramet}
\een
The transfer matrix (\ref{tra}) is the object that can be diagonalized by means of the modified algebraic Bethe ansatz \cite{B15,BelPim15,ABGP15} which extend
the usual boundary algebraic Bethe ansatz \cite{Skly88,gauge}.
\\


\paragraph{Modified Bethe vectors.} Usually, in the Bethe ansatz with $U(1)$ symmetry, the eigenvectors of the transfer matrix are generated using the operators
$\mathscr{B}(u)$ or $\mathscr{C}(u)$. In the general case, one needs modified
creation and annihilation operators (see \cite{ABGP15}) given by
\ben\label{modB}
\mathscr{B}(u,m)&=&\frac{ q u }{\gamma_{m+1}}\Big(\mathscr{B}(u)+\beta q^m\Big(\frac{q u \left(u^2-u^{-2}\right) }{\left(q u^2-q^{-1}u^{-2}\right)}\mathscr{A}(u)
-u^{-1}\mathscr{D}(u)\Big)-\beta^2 q^{2m}
\mathscr{C}(u)\Big),
\een
\ben\label{modC}
\mathscr{C}(u,m)&=&-\frac{ q u  }{\gamma_{m-1}}\Big(\mathscr{B}(u)+\alpha q^{-m} \Big(\frac{q u \left(u^2-u^{-2}\right) }{\left(q u^2-q^{-1}u^{-2}\right)}\mathscr{A}(u)
-u^{-1}\mathscr{D}(u)\Big)-\alpha^2q^{-2m}\mathscr{C}(u)\Big).
\een
Here
\ben
\gamma_{m}= \alpha
   q^{-m} -\beta q^{m}, \quad \beta=-i \frac{\tilde\kappa \tilde\xi }{\kappa \xi}q^{1-2N}, \quad
 \alpha=-i\frac{\tilde \kappa \xi}{ \kappa \tilde\xi}q^{1+2N}.
 \een
Let
\ben\label{prs}
|N\rangle = \otimes_{j=1}^N\left(\begin{array}{c}i q^{N-j}\frac{\mu\tau}{\tilde\mu\tilde\tau x_j}\\1\end{array}\right) \qquad \text{and}\qquad
\langle N| = \otimes_{j=1}^N
\left(
\begin{array}{cc}
 iq^{N-j}\frac{\mu\tilde\tau x_j}{\tilde\mu\tau} & 1 \\
\end{array}
\right)\,.
\een
Acting on these vectors with the products of the modified
creation and annihilation operators we obtain the modified Bethe vectors
\ben\label{rpsi}
|\Psi(\bar u)\rangle = \mathscr{B}(u_1,2(N-1))\mathscr{B}(u_2,2(N-2))\cdots \mathscr{B}(u_N,0)|N\rangle,
\een
and the dual ones
\ben\label{lpsi}
\langle\Psi(\bar v)| = \langle N|\mathscr{C}(v_1,2)\mathscr{C}(v_2,4)\cdots \mathscr{C}(v_N,2N).
\een
We denote  $ \bar u $ and $ \bar v $ the set of Bethe parameters $\{u_1,u_2,\dots,u_N\}$ and $\{v_1,u_2,\dots,v_N\}$. When the sets $ \bar u $ or $ \bar v $
satisfy the Bethe ansatz equations ($\mathcal{Y}(u_i|\bar u)=0$ or $\mathcal{Y}(v_i|\bar v)=0$, see (\ref{Y-Lam}) below), the vectors (\ref{rpsi},\ref{lpsi}) are called on-shell
Bethe vectors. For arbitrary
parameters  $ \bar u $ or $ \bar v $, the vectors (\ref{rpsi},\ref{lpsi}) are called off-shell Bethe vectors.


\paragraph{Off-shell equations of the transfer matrix.} Now we are in position to write down the action of the transfer matrix (\ref{tra}) on the off-shell Bethe vectors (\ref{rpsi},\ref{lpsi}), following \cite{ABGP15}. For that, it is convenient to introduce some shorthand notation, namely,
\ben
\bar u_i&=&\bar u\backslash u_i= \{u_1,\dots,u_{i-1},u_{i+1},\dots,u_N\}
\,,\quad\text{for~the~set~with~the~}u_i\text{~element~removed}\,,\non\\
\{u,\bar u_i\}&=&\{u_1,\dots,u_{i-1},u,u_{i+1},\dots,u_N\}
\,,\quad\text{for~the~set~with~the~}u_i\text{~element~replaced~by~}u\,,\non\\
Q(u,\bar u) &=& \prod_{j=1}^NQ(u,u_j) \,,\quad\text{for~the~product~of~a~two-variable~function~over~the~set~}\bar u \,,\non\\
Q(u,\bar u_i) &=& \prod_{j=1,j\neq i}^NQ(u,u_j)\,,\quad\text{for~the~product~of~a~two-variable~function~over~the~set~}
\bar u_i\,.\non
\een
If no product is involved we denote, {\it e.g.},
\ben
\Lambda(u|\bar u) &=& \Lambda(u|{u_1,\dots,u_N})\,,\non
\een
for multi-variable functions.
Using this notation, we have the following compact off-shell relations:
\ben\label{offPsi}
&&t(u)|\Psi(\bar u)\rangle  = \Lambda(u|\bar u)|\Psi(\bar u)\rangle + \sum_{i=1}^{N}\frac{F(u)}{F(u_i)}\frac{\mathcal{Y}(u_i|\bar u)}{Q(u_i,\{u,\bar u_i\})}|\Psi(\{u,\bar u_i\})\rangle,
\een
and
\ben\label{offdPsi}
&&\langle\Psi(\bar v)|t(u) =\langle\Psi(\bar v)| \Lambda(u|\bar v) + \sum_{i=1}^{N}\frac{F(u)}{F(v_i)}\frac{\mathcal{Y}(v_i|\bar v)}{Q(v_i,\{u,\bar v_i\})}\langle\Psi(\{u,\bar v_i\})|\,.
\een
Here
\ben\label{FQU}
&& F(u)=u^{-1}\frac{q^2u^2-q^{-2}u^{-2}}{q-q^{-1}}\,,\quad Q(u,v)=U(u)-U(v)\,,\quad U(u)=\frac{qu^{2}+q^{-1}u^{-2}}{(q-q^{-1})^{2}}\,,
\een
\ben\label{Lam}
&&\Lambda(u|\bar u)=\phi(u)\frac{Q(q^{-1}u,\bar u)}{Q(u,\bar u)}+\phi(q^{-1}u^{-1})\frac{Q(qu,\bar u)}{Q(u,\bar u)}+\frac{H(u)}{Q(u,\bar u)},
\een
and
\ben\label{Y-Lam}
&&\mathcal{Y}(u|\bar u)={Q(u,\bar u)}\Lambda(u|\bar u).
\een
The functions $\phi(u)$ and $H(u)$ are given by
\ben
&& \phi(u)=- \kappa \tilde \kappa \tau  \tilde\tau (\tilde \xi u+ {\tilde \xi}^{-1} u^{-1})( \xi^{-1} u+ \xi u^{-1})(\mu u+\mu^{-1} u^{-1})({\tilde \mu}^{-1}u+{\tilde \mu}u^{-1})\frac{q^2 u^2-q^{-2} u^{-2}}{q u^2-q^{-1} u^{-2}}V(u),\\
 &&
H(u) = \left( (\kappa \tau)^2+(\tilde \kappa  \tilde \tau)^2+
 \kappa \tilde \kappa \tau  \tilde\tau\left(
 \frac{\xi  \tilde \mu}{ \tilde \xi \mu} q^{N+1}+\frac{\tilde \xi \mu}{\xi \tilde \mu} q^{-N-1}\right)\right)(u^2-u^{-2})(q^2u^2-q^{-2}u^{-2})V(u) V(q^{-1}u^{-1}),
\een
where
\ben
V(u)=\prod_{i=1}^N Q(q^{1/2}u,q^{-1/2}x_i).
 \een
In the formulas above, $\Lambda(u|\bar u)$ is the eigenvalue of
the transfer matrix and $\mathcal{Y}(u|\bar u)$ is called the Bethe function, forming the Baxter TQ equation.
We can see that the Bethe function is a linear symmetric function of Bethe parameter $U(u_i)$. The structure of the eigenvalues is similar to the one of the closed XXX case but with this new variable $U(u)$, that is, the Baxter TQ equation can be written solely
in terms of $U$. It reflects the crossing invariance of the transfer matrix $t(u)=t(q^{-1}u^{-1})$. Therefore, many calculations below are similar to those in \cite{BS19det}. The differences appear in the asymptotic calculations which involve the model dependent functions $\phi(u)$ and $H(u)$.


\paragraph{Homogeneous linear system for the scalar product.} Now we can follow \cite{BS19det} to fix the general form of the scalar product of on-shell and off-shell Bethe vectors.
First of all, we consider the enlarged set $\bar u=\{u_1,\dots,u_{N+1}\}$ and define the following important products:
\ben\label{defdeltaf}
&& F(\bar u)=\prod_{u_i\in \bar u}F(u_i)\,,\quad \Delta(\bar u)=\prod_{\substack{u_i,u_j\in\bar u\\i<j}}Q(u_i,u_j)\,,\quad \Delta'(\bar u)=\prod_{\substack{u_i,u_j\in\bar u\\i>j}}Q(u_i,u_j)\,,\non\\
&&\partial U(u)= \partial_u U(u)\,,\quad \partial U(\bar u)=\prod_{u_i\in \bar u}\partial U(u_i)\,.
\een
Then we introduce variables $X_k=\langle\Psi(\bar v)|\Psi(\bar u_k)\rangle$, where $\langle\Psi(\bar v)|$ is an on-shell vector. Let us create an $N+1$ dimensional vector $X=(X_1,...,X_{N+1})$. It follows from the off-shell equations (\ref{offPsi},\ref{offdPsi}) for the transfer matrices and from $\mathcal{Y}(v_i|\bar v)=0$ with $i=1,\dots,N$ (since $\langle\Psi(\bar v)|$ is on-shell) that
\ben\label{LX}
LX=0,
\een
where $L$ is the $(N+1)\times (N+1)$ matrix with entries
\ben\label{Lmatrix}
L_{kj}=\delta_{kj}\Lambda(u_j|\bar v)-\frac{F(u_k)}{F(u_j)}\frac{\mathcal{Y}(u_j|\bar u_k)}{Q(u_j,\bar u_j)}.
\een
The homogeneous system (\ref{LX}) has non-trivial solution if $\det_{N+1}(L)=0$, and we now prove that this is indeed the case for (\ref{Lmatrix}). Recall that $\det_{N+1}(L)=0$ implies that $\text{rank}(L)\leq N$.

In order to prove that $\det_{N+1}(L)=0$, let us introduce a nonsingular $(N+1)\times (N+1)$ matrix $W$
with entries
\ben
W_{ik}=\frac{Q(u_k, \bar w_i)}{F(u_k)Q(u_k, \bar u_k)},
\een
where $\bar w=\{w_1,\dots,w_{N+1}\}$ are generic pairwise distinct parameters. The determinant of $W$ is given by
\ben
\det_{N+1}(W)= \frac{1}{F(\bar u)}\frac{\Delta(\bar w)}{\Delta(\bar u)},
\een
The product $\Omega=WL$ has the following entries:
\ben\label{Omega-1}
\Omega_{ij}&=&\frac{1}{F(u_j)Q(u_j, \bar u_j)}\Big(\Lambda(u_j|\bar v)Q(u_j, \bar w_i)-\sum_{k=1}^{N+1}\frac{Q(u_k, \bar w_i)}{Q(u_k, \bar u_k)}\mathcal{Y}(u_j|\bar u_k)\Big).
\een
To calculate the sum in \eqref{Omega-1}, we use identities
\ben\label{ident-1}
&&\sum_{k=1}^{N+1}\frac{Q(u_k, \bar w_i)}{Q(u_k, \bar u_k)}Q(a u_j,\bar u_k)=Q(a u_j,\bar w_i)\,,\quad\quad \sum_{k=1}^{N+1}\frac{Q(u_k, \bar w_i)}{Q(u_k, \bar u_k)}=1\,,
\een
where $a$ is an arbitrary complex number.  Setting $a=q$ and $a=q^{-1}$ in \eqref{ident-1} and using \eqref{Y-Lam}, we obtain
\ben
\Omega_{ij}&=&\frac{1}{F(u_j)Q(u_j, \bar u_j)}\Big(\Lambda(u_j|\bar v)Q(u_j, \bar w_i)-\mathcal{Y}(u_j|\bar w_i)\Big).
\een
Setting $\bar w_{N+1}=\bar v$ we find that the row $\Omega_{N+1,j}$ vanishes. Thus, we have $\det_{N+1}(\Omega)=0$. Therefore, it follows that  $\det_{N+1}(W)\det_{N+1}(L)=0$ and $\det_{N+1}(L)=0$.

We now find non-trivial solutions to the system (\ref{LX}). Note that the rows of $\Omega_{ij}$ with $i\leq N$ are given by
\ben
\Omega_{ij}&=&\frac{1}{F(u_j)Q(u_j, \bar u_j)}\Big(\mathcal{Y}(u_j|\bar v)\frac{Q(u_j, w)}{Q(u_j,v_i)}-\mathcal{Y}(u_j|\{\bar v_i,w\})\Big),
\een
with $w=w_{N+1}$. It is a simple exercise to show that $\Omega_{ij}$ can be rewritten as
\ben \label{Id-det}
\Omega_{ij}=\frac{Q(v_i,w)}{F(u_j)Q(u_j, \bar u_j)Q(u_j,v_i)}\mathcal{Y}(u_j|\{\bar v_i,u_j\})=\frac{Q(v_i,w)}{\partial U(v_i)}\frac{Q(u_j,\bar v)}{F(u_j)Q(u_j, \bar u_j)}\partial_{v_i} \Lambda(u_j|\bar v)\,.
\een

Now define $\tilde \Omega=V\Omega$ where $V$ is a $(N+1)\times (N+1)$  diagonal matrix with entries $V_{ii}= \frac{(1-\delta_{N+1,i})}{Q(v_i,w)}$. We have,
\ben\label{defomegatilde}
\tilde \Omega_{ij}=\frac{M_{ij}}{\partial U(v_i)F(u_j)Q(u_j, \bar u_j)}, \quad M_{ij}=Q(u_j,\bar v)\partial_{v_i} \Lambda(u_j|\bar v),
\een
The system (\ref{LX}) is equivalent to $\tilde\Omega X=0$. Let us assume that the rank of $\tilde \Omega$ is $N$ and consider the reduced matrix $\tilde \Omega_l=\tilde \Omega_{(N+1),(l)}$ which is obtained from $\tilde\Omega$ by removing the row $N+1$ and the column $l$. Then it follows from linear algebra \cite{Wh} that the system is
solved in terms of cofactors
\ben
X_l=\tilde G(\bar u| \bar v) (-1)^{N+1+l}\det_N(\tilde \Omega_l)\,,
\een
where $\tilde G(\bar u| \bar v)$ is an arbitrary function of $\bar u=\{u_1,\dots u_{N+1}\}$ and $\bar v=\{v_1,\dots,v_N\}$. Defining similarly the reduced matrix $M_l=M_{(N+1),(l)}$ for the matrix $M$ in (\ref{defomegatilde}) with the removed row $N+1$ and column $l$,  we have
\ben
X_l&=&\tilde G(\bar u| \bar v)\frac{(-1)^{N+1+l}\det_N(M_l)}
{\partial U(\bar v)F(\bar u_l)\prod_{j=1,j\neq l}^{N+1}Q(u_j, \bar u_{j})}
=\tilde G(\bar u| \bar v)\frac{(-1)^{N+1+l}\det_N(M_l)}
{\partial U(\bar v)F(\bar u_l)Q(\bar u_l,u_l)\Delta(\bar u_l)\Delta'(\bar u_l)}\non\\
&=&\frac{(-1)^N\tilde G(\bar u| \bar v)}{\partial U(\bar v)\Delta'(\bar u)}\frac{\det_N(M_l)}
{F(\bar u_l)\Delta(\bar u_l)}\,.\label{auxXl}
\een
The equation (\ref{auxXl}) implies that the ratio
\ben
\frac{F(\bar u_l)\Delta(\bar u_l)X_l}{\det_N( M_l)}=\frac{F(\bar u_j)\Delta(\bar u_j)X_j}{\det_N(M_j)}
\een
for any $l,j\in\{1,\dots,N+1\}$ is independent of $\bar u$. Therefore we can write down the solution as
\ben \label{SPwithG}
X_l=G(\bar v)\frac{\det_{N}(M_l)}{\partial U(\bar v)F(\bar u_l)\Delta(\bar u_l)},
\een
with $G(\bar v)$ an antisymmetric function of the set $\bar v$. Fixing the parameters $\bar u$ to a point where we can calculate the ``physical" scalar product allows to fix this function. For modified Bethe ansatz models we can calculate the asymptotic behavior of the scalar product for large $|u_j|$ to find it.
\\

\paragraph{Asymptotics of the determinant.}
For simplicity let us redefine  $\bar u=\bar u_{N+1}$.
To take the limit $u_j\to\infty$, $j=1,\dots,N$, we have to perform again a transformation of the determinant in (\ref{SPwithG}). First, using (\ref{Id-det}) we can rewrite it as
\ben\label{detmn1}
\det_{N}(M_{N+1})=\det_{N}\left(Q(u_j,\bar v)\partial_{v_i} \Lambda(u_j|\bar v)\right)=\partial U(\bar v)\det_{N}\left(\frac{\mathcal{Y}(u_j|\{\bar v_i,u_j\})}{Q(u_j,v_i)}\right).
\een

Let us consider $N\times N$ matrix $B$ with entries and determinant given by
\ben
B_{kl}=\frac{Q(\bar u, v_l)}{Q(u_k,v_l)Q(\bar v_l,v_l)}
\,,\quad
\det_{N}(B)=\frac{\Delta'(\bar u)}{\Delta'(\bar v)}\,.
\een
Using the matrix $B$ we can rewrite the determinant (\ref{detmn1}) as
\ben\label{detmn2}
\det_{N}(M_{N+1})=\frac{\det_{N}(BM_{N+1})}{\text{Det}_{N}(B)}=\partial U(\bar v)\frac{\Delta'(\bar v)}{\Delta'(\bar u)}\det_{N}\left(\sum_{l=1}^N\frac{\mathcal{Y}(u_j|\{\bar v_l,u_j\})}{Q(u_j,v_l)Q(u_k,v_l)}\frac{Q(\bar u, v_l)}{Q(\bar v_l,v_l)}\right).
\een
The summation in (\ref{detmn2}) is completely analogous to the XXX case considered in \cite{BS19det}. It is based on the fact that the Bethe function (\ref{Y-Lam}) is a symmetric polynomial of the first degree in the variables $U(u_i)$. One has to consider the chain rule $\partial_{U}=\frac{\partial_{u}}{\partial U(u)}$ to find that
\ben
\det_{N}(M_{N+1})=\partial U(\bar v)\Delta'(\bar v)\Delta(\bar u)\det_{N}\left(\delta_{ij}\Lambda(u_i|\bar v)+\frac{\partial_{u_j}\mathcal{Y}(w|\bar u)|_{w=u_i}}{Q(u_i,\bar u_i)\partial U(u_j)}\right).
\een
Therefore,
\ben
\frac{\det_{N}(M_{N+1})}{\partial U(\bar v)\Delta(\bar u)F(\bar u)}=\frac{\Delta'(\bar v)}{F(\bar u)}\det_{N}\left(\delta_{ij}\Lambda(u_i|\bar v)+\frac{\partial_{u_j}\mathcal{Y}(w|\bar u)|_{w=u_i}}{Q(u_i,\bar u_i)\partial U(u_j)}\right).
\een

Now the limit of the determinant at $u_i\to\infty$ can be easily calculated. To do this, we consider the asymptotic behavior of the matrix elements. First we put $u_i=u a^i$ for $i=1,\dots,N$ in (\ref{Lam}), with $a$ some generic parameter and then consider the leading term in $u$. A straightforward calculation gives
\ben\label{asylam}
\Lambda(u_i,\bar v)= \frac{(q (u a^i)^2)^{N+2}}{(q-q^{-1})^{2N}}(\tilde \kappa^2\tilde \tau^2  +\kappa^2 \tau^2 )+\cdots,
\een
and
\ben\label{asyy}
\frac{\partial_{u_j}\mathcal{Y}(w|\bar u)|_{w=u_i}}{Q(u_i,\bar u_i)\partial U(u_j)}=\frac{( q (u a^i)^2)^{N+2}}{(q-q^{-1})^{2N}}\kappa\tilde \kappa  \tau \tilde \tau \Big(\frac{\mu \tilde \xi}{\tilde \mu \xi}\Xi^{q^{-1}}_{ij}+\frac{\tilde \mu  \xi}{ \mu \tilde\xi}\Xi^{q}_{ij}\Big)+\cdots ,
\een
with
\ben
&&\Xi^q_{ij}=\frac{\prod_{k=1,k\neq j}^N(q a^{2i}-q^{-1} a^{2k})}{\prod_{k=1,k\neq i}^N( a^{2i}- a^{2k})}.
\een
Here and further on, the ellipsis denotes subleading terms compared to the leading asymptotics at $u\to\infty$.

Using (\ref{asylam},\ref{asyy}) we obtain the following asymptotic behavior of the determinant:
\ben\label{asy5}
\frac{\det_{N}(M_{N+1})}{\partial U(\bar v)\Delta(\bar u)F(\bar u)}=\frac{q^{N^2}(u^N a^{\frac{N(N+1)}{2}})^{2N+3}\Delta'(\bar v)}{(q-q^{-1})^{2N^2-N}}\nu_N+\cdots\,,
\een
with
\ben
\nu_N&=&\det_{N}\left((\tilde \kappa^2\tilde \tau^2  +\kappa^2 \tau^2 )\delta_{ij}+\kappa\tilde \kappa  \tau \tilde \tau \Big(\frac{\mu \tilde \xi}{\tilde \mu \xi}\Xi^{q^{-1}}_{ij}+\frac{\tilde \mu  \xi}{ \mu \tilde\xi}\Xi^{q}_{ij}\Big)\right)\non\\
&=&(\kappa^2 \tau^2 )^N\Big(-\frac{\tilde\kappa \tilde\tau \tilde \mu \xi}{\kappa \tau \mu \tilde  \xi}q^{1-N};q^2\Big)_N \Big(-\frac{\tilde\kappa \tilde\tau \mu  \tilde \xi}{\kappa \tau \tilde \mu   \xi}q^{1-N};q^2\Big)_N.
\een
Here
\ben
(b;q)_n=\prod_{k=0}^{n-1}(1-bq^{k})
\een
is the q-Pochhammer symbol.
\\

\paragraph{Asymptotics of the Bethe vectors}
The next step is to compute the asymptotic behavior of the scalar product $\langle\Psi(\bar v)|\Psi(\bar u)\rangle$ through the objects of the quantum inverse scattering method.
Using the R-matrix (\ref{Rmatrix}) we obtain for $u\rightarrow\infty$
\ben\label{Tasy}
&&\left(\prod_{i=1}^Nx_i\right) T_a(u)  =\left(\frac{q u^2}{(q-q^{-1})^2}\right)^{N/2}\left(
\begin{array}{cc}
q^{S^3}   & 0 \\
0 &  q^{-S^3}   \\
\end{array}
\right)+ \left(\frac{q u^2}{(q-q^{-1})^2}\right)^{(N-1)/2}\left(
\begin{array}{cc}
0 & S^{-} \\
S^{+} &  0  \\
\end{array}
\right)+\cdots,
\een
and
\ben\label{Thasy}
&&\left(\prod_{i=1}^Nx_i\right)^{-1}\hat T_a(u)  =\left(\frac{q u^2}{(q-q^{-1})^2}\right)^{N/2}\left(
\begin{array}{cc}
q^{S^3}   & 0 \\
0 &  q^{-S^3}   \\
\end{array}
\right)+\left(\frac{q u^2}{(q-q^{-1})^2}\right)^{(N-1)/2}\left(
\begin{array}{cc}
0 & \hat S^{-} \\
\hat S^{+} &  0  \\
\end{array}
\right)+\cdots .
\een
The $U_q(sl_2)$ operators in (\ref{Tasy},\ref{Thasy}) are
\ben
S^3=\sum_{i=1}^N \frac{\sigma^{3}_i}{2},
\quad\quad
S^\pm=
\sum_{i=1}^N x_i q^{\mp\sum_{j=1}^{i-1} \frac{\sigma^3_j}{2}}\sigma^{\pm}_i q^{\pm\sum_{j=i+1}^{N}  \frac{\sigma^3_j}{2}},
\een
and
\ben
\hat S^\pm=\sum_{i=1}^N x_i^{-1} q^{\pm\sum_{j=1}^{i-1} \frac{\sigma^3_j}{2}}\sigma^{\pm}_i q^{\mp\sum_{j=i+1}^{N}  \frac{\sigma^3_j}{2}}.
\een

Substituting (\ref{Tasy},\ref{Thasy}) in (\ref{KO}) we obtain for $u\rightarrow\infty$
\ben\label{Kaasy}
K_a(u) &=&\left(\frac{q u^2}{(q-q^{-1})^2}\right)^{N} \left(
\begin{array}{cc}
A u  & \tau^2 u^2 \\
\tilde\tau^2 u^2 &   A^* u  \\
\end{array}
\right)
+\cdots\,,
\een
where
\ben
A&=&\nu_- q^{2 S^3}+(q-q^{-1})q^{-1/2}(\tilde\tau^2S^-q^{S^3}+\tau^2 q^{S^3} \hat S^+),
\\
A^*&=&\nu_+ q^{-2 S^3}+(q-q^{-1})q^{-1/2}(\tau^2S^+q^{-S^3}
+\tilde\tau^2 q^{-S^3} \hat S^-),
\een
that satisfy the q-deformed Dolan-Grady relations \cite{B04},
\ben
\big[A,\big[A,\big[A,A^*\big]_q\big]_{q^{-1}}\big]&=&(q^2-q^{-2})^2\tau^2\tilde\tau^2\big[A,A^*\big]\ ,\nonumber\\
\big[A^*,\big[A^*,\big[A^*,A\big]_q\big]_{q^{-1}}\big]&=&(q^2-q^{-2})^2\tau^2\tilde\tau^2\big[A^*,A\big]\,,
\een
where $[A,B]_q=qAB-q^{-1}BA$.

It follows from (\ref{Kaasy}) that the modified operator (\ref{modB}) behaves as
\ben
&&\mathscr{B}(u,m)=\left(\frac{q u^2}{(q-q^{-1})^{2}}\right)^{N}q u^3 \frac{\beta q^{m}}{\gamma_{m+1}} \left(A
+q^{-m}\beta^{-1}\tau^2
-q^m
\beta \tilde\tau^2\right)+\cdots\,,
\een
as $u\rightarrow\infty$.
\\

The  action of $A$ on the  vector (\ref{prs}) follows from the representation theory of the
q-Onsager algebra \cite{B06}. In our notation, the action of $A$ is
\ben
A|N\rangle = i\tau\tilde\tau\left(q^{N}\frac{\mu}{\tilde\mu}+q^{-N}\frac{\tilde\mu}{\mu}\right)|N\rangle.
\een
Therefore,
\ben\label{asybb}
\mathscr{B}(u,m)|N\rangle = \left(\frac{q u^2}{(q-q^{-1})^{2}}\right)^{N} q u^3 \tau^2
\frac{
\left(1+i\frac{\tilde\mu\tilde\tau}{\mu\tau}q^{-N+m}\beta \right)\left(1+i\frac{\mu\tilde\tau}{\tilde\mu\tau}q^{N+m}\beta\right)}{\gamma_{m+1}}
|N\rangle+\cdots\,,
\een
as $u\rightarrow\infty$. Formulas (\ref{asybb}) and (\ref{rpsi}) imply that the scalar product $\langle\Psi(\bar v)|\Psi(\bar u)\rangle$ has the following asymptotics:
\ben\label{asySP}
&&\langle\Psi(\bar v)|\Psi(\bar u)\rangle=\frac{(\bar u)^{2N+3}\langle\Psi(\bar v)|N\rangle}{(q-q^{-1})^{2 N^2}}\left(i\frac{\kappa \tilde\xi}{\tilde\kappa \xi}\tau^{2}\right)^N\frac{\left(-\frac{\tilde\kappa
   \tilde\tau \tilde\xi \tilde\mu }{\kappa \tau \xi  \mu  }q^{1-3 N};q^2\right)_N \left(-\frac{\tilde\kappa  \tilde\tau  \mu  \tilde\xi}{\kappa  \tau
   \xi  \tilde\mu}q^{1-N};q^2\right)_N}{\left(\frac{ \tilde\xi^2}{\xi ^2}q^{2-4 N};q^4\right)_N}+\cdots\,,
\een
for $u_i\rightarrow\infty$.
\\


\paragraph{Determinant formula} We are in position to present
the scalar product of the on-shell dual Bethe vectors $\langle\Psi(\bar v)|$ and off-shell Bethe vectors $|\Psi(\bar u)\rangle$. To do this, it is enough to compare \eqref{asySP} with the formula \eqref{SPwithG} for $u_i\to\infty$ (see (\ref{asy5})). This allows us to find the unknown function $G(\bar v)$. A simple calculation shows that
\ben\label{scalar}
\langle\Psi(\bar v)|\Psi(\bar u)\rangle =\eta\langle\Psi(\bar v)|N\rangle  \frac{\det_{N}
\left(Q(u_j,\bar v)\partial_{v_i} \Lambda(u_j|\bar v)\right)}{\partial U(\bar v)\Delta'(\bar v)F(\bar u)\Delta(\bar u)},
\een
where
\ben
&&\eta= \left(\frac{i \tilde\xi}{q^N(q-q^{-1})\tilde\kappa \kappa \xi}\right)^N
\frac{\left(-\frac{\tilde\kappa \tilde\tau \tilde\mu  \tilde\xi}{\kappa   \tau  \mu \xi }q^{1-3N};q^2\right)_N}
{\left(\frac{ \tilde\xi^2}{\xi ^2}q^{2-4 N};q^4\right)_N\Big(-\frac{\tilde\kappa \tilde\tau \tilde \mu \xi}{\kappa \tau \mu \tilde  \xi}q^{1-N};q^2\Big)_N}\,.
\een
This is the main result of the paper. Recall that the functions entering (\ref{scalar}) are given in (\ref{FQU},\ref{Lam},\ref{Y-Lam},\ref{defdeltaf}).
\\

\paragraph{XXX limit} We can now check the conjecture of \cite{BP15scalar2} for the XXX chain. We need to consider the change of variables
\ben
&&u=e^{\hbar \lambda}, \quad v=e^{\hbar \upsilon}, \quad x=e^{\hbar \theta}, \quad q= e^{\hbar},  \quad  \epsilon_\pm=\pm\frac{e^{\pm\hbar \mathfrak{q}}}{ e^{\hbar}- e^{-\hbar}},\quad \nu_\pm=\mp\frac{e^{\mp\hbar p}}{ e^{\hbar}- e^{-\hbar}}, \quad  \frac{\tilde\xi}{\xi}=i\frac{\rho}{\sqrt{\xi^+\xi^-}}\,,\nonumber
 \\
&& \kappa^2=\frac{\xi^-}{2( e^{\hbar}- e^{-\hbar})},  \quad \tilde \kappa^2=\frac{\xi^+}{2( e^{\hbar}- e^{-\hbar})}
,  \quad \tau^2=\frac{\eta^+}{2( e^{\hbar}- e^{-\hbar})},  \quad \tilde\tau^2=\frac{\eta^-}{2( e^{\hbar}- e^{-\hbar})},  \quad  \frac{\mu}{\tilde \mu}=i\frac{\sqrt{\eta^+\eta^-}}{\tilde \rho}\,,
\nonumber
\een
and take the limit\, $\hbar \to 0$. We find for the limit of the eigenvalue
\ben\label{LamX}
&&\Lambda_X(\lambda |\bar \lambda)=\phi_X(\lambda)\frac{Q_X(\lambda-1,\bar \lambda)}{Q_X(\lambda,\bar \lambda)}+\phi_X(-\lambda-1)\frac{Q_X(\lambda+1,\bar \lambda)}{Q_X(\lambda,\bar \lambda)}+\frac{H_X(\lambda)}{Q_X(\lambda,\bar \lambda)},
\een
where
\ben
Q_X(\lambda,\theta)=U_X(\lambda)-U_X(\theta), \quad U_X(\lambda)=\lambda(\lambda+1), \quad \partial U_X(\lambda)=2\lambda+1,
\een
\ben
&& \phi_X(\lambda )=(\mathfrak{q}+ \lambda(1-\rho))(p+ \lambda(1-\tilde \rho))V_X(\lambda )\frac{2\lambda +2}{2\lambda +1},\\
 &&
H_X(u) =\left(\xi^- \eta^++ \xi^+ \eta^-+2-
 2(\rho -1)(\tilde\rho-1) \right) \lambda (\lambda+1)V_X(\lambda )V_X(-\lambda -1),\\
 &&
V_X(\lambda)=\prod_{i=1}^N Q(\lambda+1/2,\theta_i-1/2), \quad F_X(\lambda)=2\lambda+2.
 \een
Here  $(1-\tilde \rho)^2 =1+\eta^+\eta^-$ and $(1-\rho)^2=1+\xi^+\xi^-$.

Then we obtain for the scalar product
 \ben\label{scalarX}
\langle\Psi_X(\bar  \upsilon)|\Psi_X(\bar \lambda)\rangle =
\Big(\frac{(\rho-2)}{2( \rho-1)^2}{\frac{\eta^+\eta^-}{\tilde\rho^2}}\frac{1+\frac{\tilde\rho\rho}{\eta^+\xi^-}}{1-\frac{\tilde\rho\xi^+}{\eta^+\rho}}\Big)^N\langle X|\tilde{\mathscr{C}}(\bar \upsilon) |X\rangle  \frac{\det_{N}
\left(Q_X(\lambda_j,\bar  \upsilon)\partial_{ \upsilon_i} \Lambda_X(\lambda_j|\bar  \upsilon)\right)}{\partial U_X(\bar  \upsilon)\Delta_X'(\bar  \upsilon)F_X(\bar \lambda)\Delta_X(\bar \lambda)},
\een
where we used $\xi^+\xi^-=\rho(\rho-2)$. The operator $\tilde{\mathscr{C}}(\upsilon)$ and the vectors $|X\rangle$ and $\langle X|$ are given by
\ben
&&\tilde{\mathscr{C}}(\upsilon)=\frac{\xi^+}{\rho}\mathscr{C}_X(\upsilon)+\Big(\frac{2\upsilon }{2\upsilon+1}\mathscr{A}_X(\upsilon)
-\mathscr{D}_X(\upsilon)\Big)-\frac{\rho}{\xi^+}\mathscr{B}_X(\upsilon), \\
&&|X\rangle = \otimes_{j=1}^N\left( \begin{array}{c}1\\{-\frac{\tilde \rho}{\eta^+}}\end{array}\right)\,,\quad
\langle X| = \otimes_{j=1}^N
\left(
\begin{array}{cc}
1 & {-\frac{\tilde\rho}{\eta^-} }\\
\end{array}
\right)\,.
\een
Taking the limit $\tilde\rho\to 0$ and then $\eta^\pm\to 0$ of the ratio $\langle\Psi_X(\bar  \upsilon)|\Psi_X(\bar \lambda)\rangle /\Big(\frac{\eta^+\eta^-}{\tilde\rho^2}\Big)^N$ we recover the conjecture of \cite{BP15scalar2} and thus prove it.
\\


\paragraph{Conclusion.} In this paper, we computed
the scalar product of on-shell and off-shell modified Bethe vectors
associated with the XXZ spin chain with general integrable boundary conditions.
Similarly to the $U(1)$-invariant models, the compact formula is
given by a determinant involving the Jacobian of the transfer
matrix eigenvalue. Breaking the $U(1)$ symmetry implies the expectation value $\langle\Psi(\bar v)|N\rangle$, while the ``dynamic" nature of the modified creation operators generates q-Pochhammer symbols involving the boundary parameters fields. The first point to be investigated from here is the computation of the expectation
value $\langle\Psi(\bar v)|N\rangle$ and check whether it can be written as a boundary case of the q-deformed modified Izergin determinant \cite{BS19,Tsu,FW}. The next natural step is to compute
the action of local operators on the modified Bethe vector, and then derive building
blocks of correlation functions, trying to generalize the known results for the diagonal boundary case \cite{KKMNST,KKMNST2}. Certainly, the solution of these problems will
bring advances to the solution of the inverse
problem for models without $U(1)$ symmetry, providing new tools
for the calculation of physical quantities.

\vspace{0.5cm}

\noindent{\bf Acknowledgments.}  R.P. thanks the Institute Denis Poisson,
where a part of this work was done, for hospitality, and Pascal Baseilhac
for discussions and encouragement. R.P. was supported by CNPq (grant \# 150829/2020-5).  N.S. thanks the CNRS for his grant to join the Institute Denis Poisson of the University of Tours where a part of this work was done.

\end{document}